\documentclass[12pt, a4paper]{article}
\usepackage{afterpage}
\usepackage{algorithm}
\usepackage{algorithmicx}
\usepackage{algpseudocode}
\usepackage{authblk}
\usepackage{amsmath}
\usepackage{braket} 
\usepackage[title]{appendix}
\usepackage{amsfonts}
\usepackage{booktabs} 
\usepackage[citestyle = apa, bibstyle=apa, maxbibnames=20]{biblatex}
\usepackage[font=small,labelfont=bf, hypcap=false]{caption} 
\usepackage{chngcntr}
\usepackage{comment} 
\usepackage{csquotes} 
\usepackage{csvsimple} 
\usepackage{diagbox}
\usepackage{eqnarray}
\usepackage{float} 
\usepackage[T1]{fontenc} 
\usepackage{geometry}
\usepackage{graphicx}
\usepackage[colorlinks=true, allcolors=blue]{hyperref}
\usepackage{lineno} 
\usepackage{listings}
\usepackage{lipsum}
\usepackage{longtable}
\usepackage{lscape}
\usepackage{mathrsfs}
\usepackage{multirow}
\usepackage{orcidlink}
\usepackage{pdflscape}
\usepackage[]{ragged2e}
\usepackage{subcaption}
\usepackage{setspace} 
\usepackage{threeparttable} 
\usepackage{tabularray}
\usepackage{tabularx}
\usepackage{tcolorbox} 
\usepackage{titlesec} 
\usepackage{xcolor} 
\usepackage{dirtree}
\usepackage{markdown}

\lstdefinelanguage{JSON}{
    morekeywords={true,false,null},
    sensitive=false,
    morecomment=[l]{//},
    morecomment=[s]{/*}{*/},
    morestring=[b]",
    alsoletter={-},
    stringstyle=\color{red},
    keywordstyle=\color{blue},
    commentstyle=\color{gray}\itshape
}

\lstdefinelanguage{PlainText}{
    keywords={},
    sensitive=false,
    comment=[l]{//},
    commentstyle=\color{gray}\itshape,
    morestring=[b]"
}

\newcommand{\code}[1]{%
    \colorbox{gray!10}{\ttfamily\detokenize{#1}}%
}

\titleformat{\section} 
{\normalfont\Large\bfseries}{\thesection.}{1em}{}

\newcommand{\subsubsubsection}[1]{%
  \vspace{\baselineskip} 
  \noindent\textbf{#1\\}\quad 
}



\title{\textbf{MiqroForge: An Intelligent Workflow Platform for Quantum‑Enhanced Computational Chemistry}}

\author[1,2]{\small Jianan Wang}
\author[3]{\small Wenbo Guo}
\author[3]{\small Xin Yue}
\author[2]{\small Minjie Xu}
\author[2]{\small Yueqiang Zheng}
\author[3]{\small Jingxiang Dong}
\author[3]{\small Jiarui Hu}
\author[2]{\small Jian Xia}
\author[1,2,*]{\small Chuixiong Wu}

\affil[1]{\small{Suzhou Miqro Era Quantum Technology Co. Ltd., Suzhou, China}}
\affil[2]{\small{Shanghai Miqro Era Digital Technology Co. Ltd., Shanghai, China}}
\affil[3]{\small{Hefei Miqro Era Digital Technology Co. Ltd., Hefei, China}}
\affil[*]{Corresponding author: \href{mailto:wuchuixiong@miqroera.com}{wuchuixiong@miqroera.com}}
\date{}

\begin{document}

\newpage
\maketitle
\begin{abstract}
{The connect-fill-run workflow paradigm, widely adopted in mature software engineering, accelerates collaborative development. However, computational chemistry, computational materials science, and computational biology face persistent demands for multi-scale simulations constrained by simplistic platform designs. We present MiqroForge, an intelligent cross-scale platform integrating quantum computing capabilities. By combining AI-driven dynamic resource scheduling with an intuitive visual interface, MiqroForge significantly lowers entry barriers while optimizing computational efficiency. The platform fosters a collaborative ecosystem through shared node libraries and data repositories, thereby bridging practitioners across classical and quantum computational domains.}
\end{abstract}

\\
\noindent\textbf{Github Repository}: \url{https://github.com/MiqroEra/MiqroForge} \\
\noindent\textbf{Online Documentation}: \url{https://miqroforge-docs.readthedocs.io} \\

\newpage

\tableofcontents

\newpage

\section{Introduction}

In the field of modern engineering and scientific research, workflow management has become the core paradigm of complex system management due to its structured task scheduling and resource optimization capabilities. Its applications are accelerating, extending from traditional software engineering to high-performance computing, scientific simulation, and creative industries. In AI creation content, especially in the field of image generation, platforms represented by Comfy UI\footnote{https://github.com/comfyanonymous/ComfyUI} have greatly improved the application of Diffusion models(\cite{podell2023sdxlimprovinglatentdiffusion}). Similarly, in scientific computing, platforms like Taverna\footnote{https://incubator.apache.org/projects/taverna} demonstrate workflow efficacy in bioinformatics, cheminformatics, medicine, astronomy, social science, music, and digital preservation.

Modern molecular computational simulation, such as computational chemistry, computational materials, computational biology, and other fields, has long faced problems such as frequently updating algorithms, lengthy workflows, and difficult resource management. It is difficult for developers to conduct application-level testing, and it is inconvenient for engineers to call the latest algorithms. This creates a waste of resources and bottlenecks in the industry. Existing solutions such as AiiDA(\cite{PIZZI2016218}), Fireworks(\cite{jain2015fireworks}), Cuby(\cite{vrezavc2016cuby}), etc., have achieved a lot of acceptance, but still suffer from several critical limitations: There is a lack of further standardization of applications, so these tools are often hosted as platforms rather than ecosystems, and users can basically only refactor existing source code, and it is not convenient to use these tools to expand the scope of research; Some platforms are organized in a single programming language (e.g., Python only) and therefore sometimes lack compatibility for cross-language applications; often confined to a certain area, such as high-throughput screening; The overall computing resource scheduling is not intelligent enough; Lacks a user-friendly interface.


On the other hand, the application of quantum computing in the field of chemistry has recently received extensive attention(\cite{duriez2025computing}). Traditional computational chemistry is difficult to realize value in the application areas currently being discussed because it does not address the problem of strong correlation, which may be one of the reasons why workflow platforms have not yet received attention in computational chemistry - if computational simulation is always seen as an add-on, there will be insufficient incentive to develop its ecosystem.Once mixtures of classical and quantum algorithms (e.g., VQE \cite{peruzzo2014variational}, QSCI \cite{kanno2023quantum}, etc.) can provide better predictions for actual systems, attempts to build workflows will increase greatly. At the same time, researchers of quantum algorithms are also looking for systems that can simplify the quantum computing process and improve application connectivity(\cite{alexeev2025perspective}). While platforms like CUDA-Q(\cite{kim2023cuda}) provide quantum programming frameworks, they remain inaccessible to non-specialists due to code-centric implementations. From another point of view, beyond traditional QM/MM, practical workflows often embed quantum algorithms (e.g., active-space methods) atop HF/DFT layers (e.g., DMET(\cite{knizia2013density}) or DDA(\cite{gujarati2023quantum})) while interfacing with force-field scale samplers, forming a triple‑embedding pattern. This motivates workflow‑first design that standardizes I/O and resource decisions across scales. 

Overall, while there has been some exploration in using workflows to manage computational chemistry processes, the full process has not yet been addressed. MiqroForge addresses these challenges through: 1. A node-centric architecture enabling workflow reuse and community-driven development. 2. An intuitive visual interface. 3. AI-optimized resource allocation dynamically balancing HPC and quantum resources. 4. Extensible quantum modules for electronic structure calculations. These are discussed in more detail in Section 2. MiqroForge is released under a dual-licensing model. The Community Edition is provided under the PolyForm Noncommercial License 1.0.0, which permits non-commercial use, modification, and distribution. Commercial use (including offering MiqroForge as part of a paid product or service) requires a separate commercial license from Miqro Era. We welcome individual contributors to develop and share nodes under the same community license; a contributor license agreement (CLA) is used to enable dual licensing. Repository and documentation links are provided for details. The platform includes quantum-chemistry workflow templates for catalytic simulations and strongly correlated systems, lowering the barrier to quantum-classical hybrid computing.

The paper proceeds as follows: \textbf{Section 2} details MiqroForge's basic idea, the architecture and installation. \textbf{Section 3} demonstrates the user interface (UI) through a catalytic reaction case study. \textbf{Section 4} defines node structures and creation protocols. \textbf{Section 5} specifies input/output core mechanisms. \textbf{Section 6} introduces AI-driven resource scheduling. \textbf{Section 7} discusses data persistence strategies and workflow governance. \textbf{Section 8} outlines the development roadmap. Consistent with our focus on architectural innovation, implementation details are minimized in favor of design principles.

\section{Platform Overview}

MiqroForge constitutes a modular, multi-layered, and intelligent multi-scale molecular design platform. Researchers familiar with quantum computing (chemistry), materials computation, quantum chemistry, AI4S, or molecular dynamics may leverage MiqroForge for algorithm development and application construction. The platform employs workflows to encapsulate computational processes across chemical, materials, and biological domains, thereby reducing development complexity, enhancing resource scheduling flexibility, and facilitating cross-domain collaboration.

\begin{center}
\includegraphics[width=\linewidth, clip] {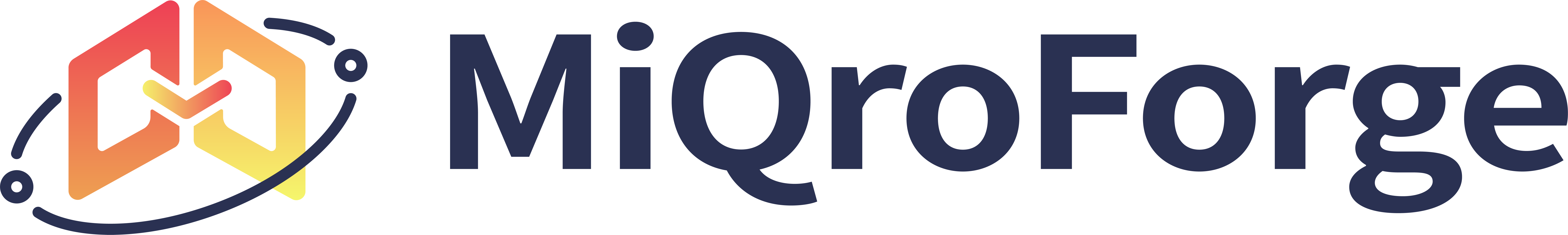}
\captionof{figure}{\small Logo of MiqroForge}
\label{Figure: logo}
\end{center}

\subsection{Features}

MiqroForge supports computations spanning electron wavefunction to molecular cluster scales, with planned extensions for experimental data synchronization. It exhibits three principal characteristics:

\begin{center}
\includegraphics[width=\linewidth, clip] {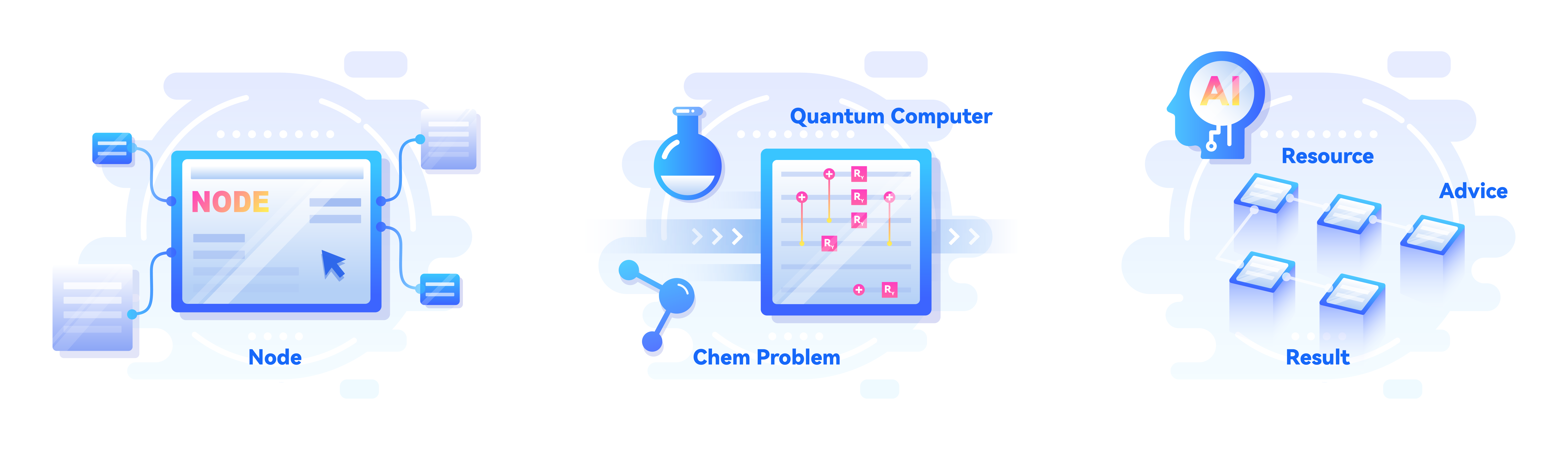}
\captionof{figure}{\small Three core features: node‑centric computation, quantum integration, and AI‑driven scheduling.}
\label{Figure: demostra}
\end{center}

\begin{enumerate}
    \item \textbf{Node-based computation}: Standardized interfaces and resource scheduling systems transform nodes beyond process steps into reusable productivity tools. Users may concentrate on single-node development and cross-scale applications without expertise in other computational scales.
    \item \textbf{Quantum computing integration}: The platform incorporates quantum computing nodes, enabling researchers to apply quantum methods to practical scenarios. Existing workflows can be migrated to this quantum-enhanced environment.
    \item \textbf{AI-driven optimization}: Unlike conventional platforms requiring manual resource allocation, MiqroForge automates scheduling. Node developers specify resource scaling parameters in \code{performance_config.json} (Section \ref{sec: p_c_json}), allowing users to initiate computations without managing underlying infrastructure.
\end{enumerate}

\subsection{Target Users}

Researchers across multiple disciplines will benefit from MiqroForge:

\begin{itemize}
    \item \textbf{Algorithm developers}: Developers can utilize MiqroForge to concentrate on algorithmic implementation and performance optimization. The platform's resource scheduling functionality enables focus on temporal and spatial resource consumption; Integrated application workflows facilitate rapid algorithm validation; Flexible node architecture supports comparative algorithm analysis; Extensive online resources provide streamlined access to cloud computing infrastructure.
    \item \textbf{Team Leaders/Mentors}: Pre-configured workflows enhance pedagogical effectiveness; Modular task delegation accelerates project execution.
    \item \textbf{Applied Researchers/Engineers}: Pre-optimized computational nodes reduce workflow configuration complexity; Integration of novel and quantum algorithms within traditional processes is enabled; Intelligent scheduling and cloud resources expedite high-throughput screening; Visualization nodes support comprehensive process documentation.
\end{itemize}

Additionally, we invite computing resource providers to participate in collaborative development of our cloud resource scheduling infrastructure and allocation strategies.

\subsection{Quantum Computational Chemistry}

Quantum computing represents both a critical component of MiqroForge and an emerging methodology with significant potential for electronic wavefunction problems.

Diverging from classical computing in hardware and computational principles, quantum computers utilize superposition states ($
(\ket{0}+\ket{1})/\sqrt{2}$) and gates (e.g., Hadamard) that surpass classical logic operations. These properties enable quantum advantage ("quantum supremacy") for specific problems, albeit with fundamentally distinct algorithmic foundations.

Notably, computational chemistry solves the second-quantized Hamiltonian:

\begin{equation}
    \hat{H}=\sum_{pq}h_{pq}a_p^\dagger a_q + \frac{1}{2}\sum_{pqrs}g_{pqrs}a_p^\dagger a_r^\dagger a_sa_q + h_{\mathrm{nuc}}
\end{equation}

yielding (ground) electronic Fock states. Within a specified basis set, quantum algorithms achieve precision comparable to Full Configurational Interaction (CI). Where Density Functional Theory (DFT) encounters limitations in capturing strong correlation effects (often addressed via DFT$+$U$+$V corrections (\cite{duriez2025computing})), quantum computing provides alternative solutions. Consequently, quantum computing interfaces with both primary computational chemistry methodologies: Wavefunction Theory (WFT) and DFT.

As an emerging computational paradigm, quantum computing demonstrates potential for addressing exponential scaling of active spaces and strong correlation challenges in computational chemistry.

\subsection{Installation} \label{sec: install}

The installation procedure commences with source code acquisition from the GitHub repository:

\begin{lstlisting}[language=bash]
git clone https://github.com/MiqroEra/MiqroForge
cd MiqroForge
\end{lstlisting}

Subsequent deployment utilizes an integrated installation script configuring Docker, Kubernetes, and Web UI components:

\begin{lstlisting}[language=bash]
bash scripts/install_miqroforge.sh
\end{lstlisting}

Note: Installation procedures correspond to the current version. For updates or unresolved issues, consult the latest documentation or GitHub repository.

Retrieving container images requires substantial time due to the Web UI services, intelligent components, and quantum chemistry nodes. Upon successful completion, initiate services via:

\begin{lstlisting}[language=bash]
miqroforge run -p 30080 -ip localhost
\end{lstlisting}

Access the Web UI at \code{http://localhost:30080}. If port 30080 is occupied, use -p <alternative\_port> and access http://localhost:<alternative\_port>. Initial deployment on Windows Subsystem for Linux (WSL) is recommended. While WSL resources are inadequate for production-scale computation, they suffice for preliminary demonstrations. Server deployment for network sharing requires additional configuration (see documentation).

Common commands include:

\begin{lstlisting}[language=bash]
miqroforge show                # Display local nodes
miqroforge status --detail     # Node-level computation status
miqroforge resources --live    # Resource utilization
miqroforge task submit simulation-job.yaml  # Workflow submission
miqroforge task list           # Task enumeration
\end{lstlisting}

Web UI usage is recommended for standard operations; command-line interfaces serve as supplementary options. Consult online documentation for implementation details.

\subsection{Architecture Framework}

Post-installation, users construct workflows using existing nodes via Web UI. However, node creation necessitates understanding of the architectural framework:

\begin{center}
\includegraphics[width=\linewidth, clip]{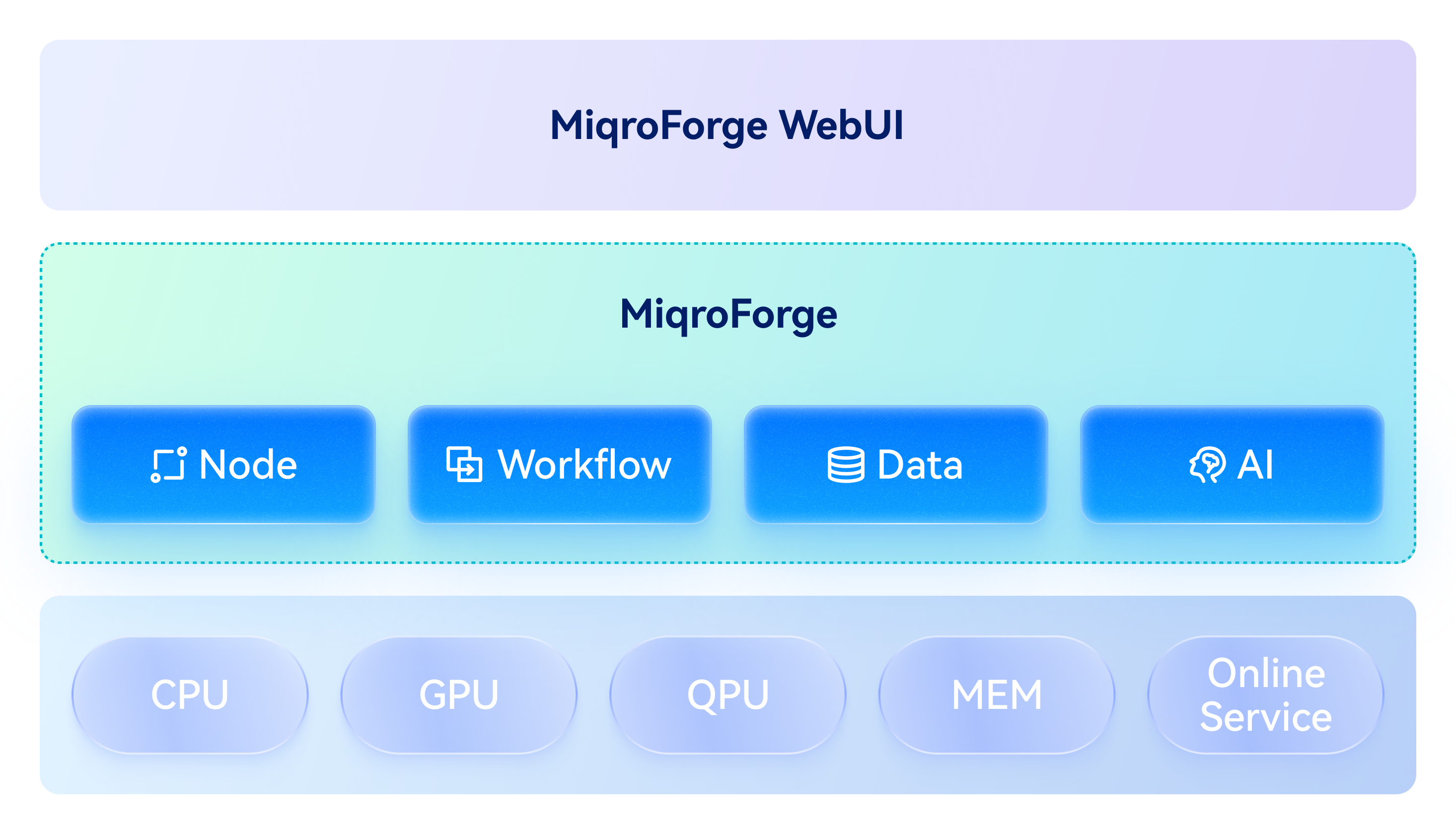}
\captionof{figure}{\small Schematic diagram of the architecture framework}
\label{Figure: arch}
\end{center}

As illustrated in Figure \ref{Figure: arch}, MiqroForge comprises four core components beyond the Web UI: \textbf{Node}, \textbf{Workflow}, \textbf{Data}, and \textbf{AI}. These elements coordinate algorithms and data with allocated computational resources (CPU/GPU) for closed-loop computation. \textbf{Nodes}—defined as executable units with singular functions and standardized I/O—constitute the fundamental modular elements. \textbf{Workflows} establish informational and computational relationships between nodes. \textbf{Data} management encompasses task-specific information and result databases, including heterogeneous node and workflow data. The \textbf{AI} component provides computational scheduling, workflow recommendations, automated reporting, and supports Agent functions through plugin interfaces.

Additional considerations:
\begin{enumerate}
    \item Quantum computing functionality operates via internal QPU calls within nodes, thus not constituting an architectural layer. Documentation details quantum node implementation, as with AI4S-based models.
    \item Future development may incorporate hardware-derived measurement data (e.g., spectrometer outputs), though this capability remains outside the current version's scope.
\end{enumerate}

\section{Website User Interface}

\subsection{Functional Area}

As introduced in Section \ref{sec: install}, after successfully launching MiqroForge, users can see the user interface by accessing the specified web address, typically \code{http://localhost:30080}.

\begin{center}
\includegraphics[width=\linewidth, clip]{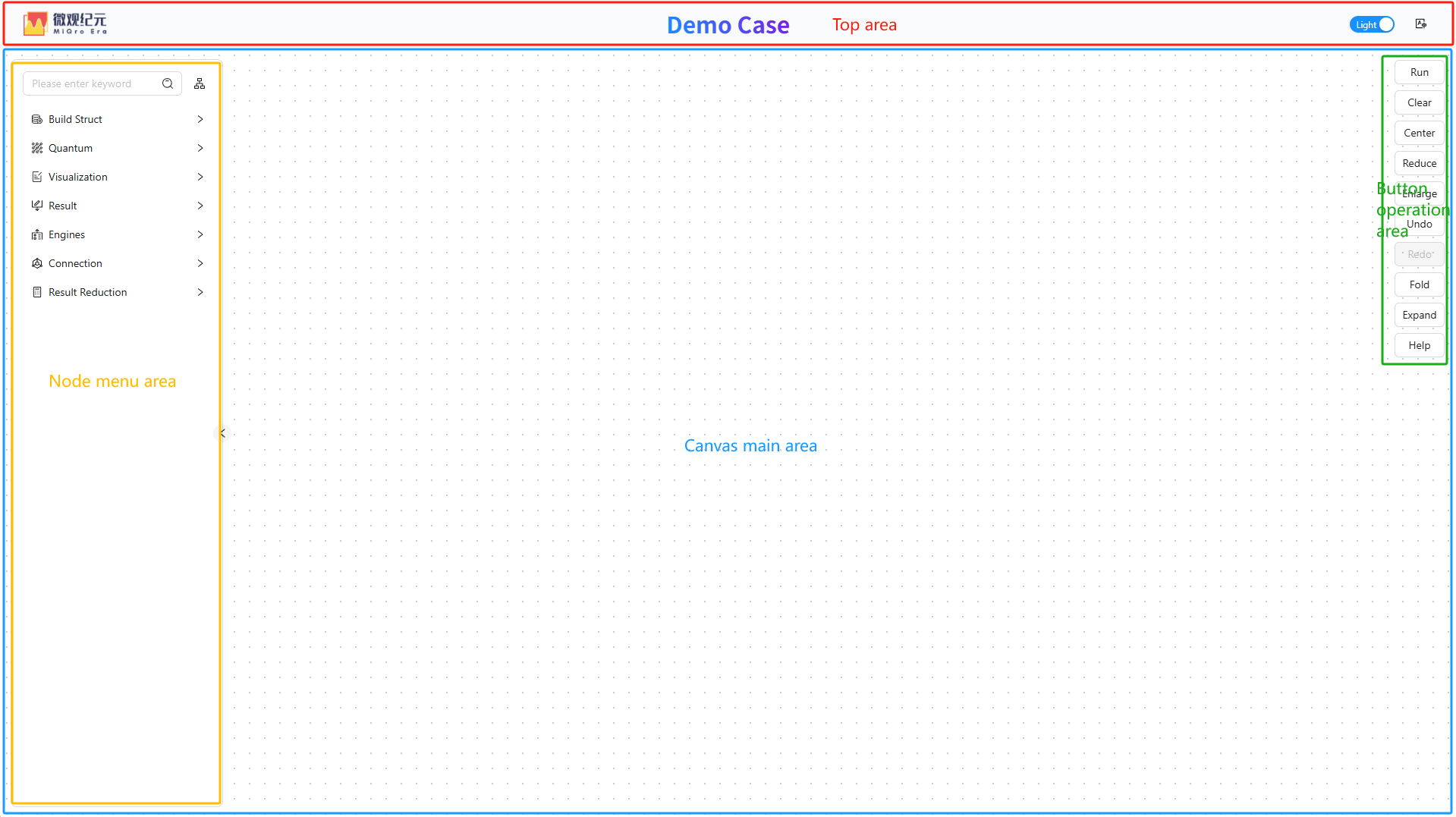}
\captionof{figure}{\small Workflow build page in the Web UI}
\label{Figure: Web UI}
\end{center}

The entire user interface is divided into four main areas. In the center is the canvas area, which is also the main area. The menus on the left and right are the node menu and the button action area, respectively. The top part is the overview area.

A typical workflow build process is to select the nodes you need in the left node area and drag them onto the center canvas. Connect nodes on the central canvas and complement the necessary start/end process, or add a logical judgment section. Once the workflow is built in the central area, fill in the required information for each node. Finally, select Run, Terminate, or Action on the canvas in the right area. The area above allows you to name the workflow and save the workflow. It is also possible to switch to other pages, such as making a new node.

\subsection{Build and Run}

Within the MiqroForge v1, there comes a classic example of calculating the potential energy surface of water molecules using the quantum computational chemistry algorithm QSCI. This workflow demonstrates MiqroForge's extensibility, as users can replace QSCI with VQE nodes for comparative analysis. Of course, since these nodes are already public, users can also use these nodes to build their own quantum computational chemistry workflows. At least, extending QSCI to other small molecules is not a problem, as long as the user has a simple understanding of how to use QSCI through the instructions on the node. This is a first demonstration of the power of MiqroForge, with standardized nodes that allow users to quickly learn and efficiently reuse workflows.

\begin{center}
\includegraphics[width=\linewidth, clip]{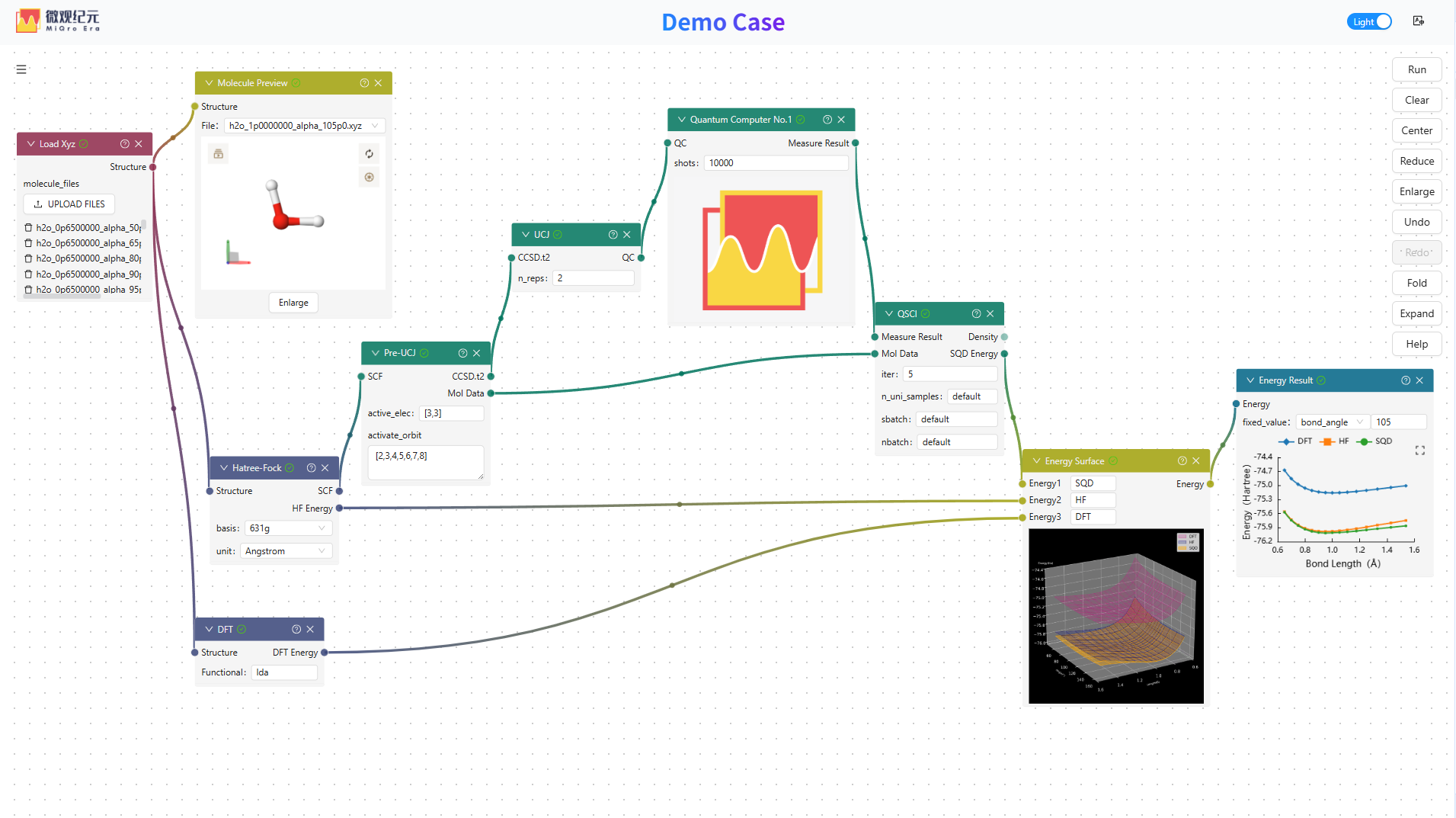}
\captionof{figure}{\small Example of calculating the potential energy surface of water molecules using the quantum computational chemistry algorithm QSCI}
\label{Figure: demo}
\end{center}

Here, we will talk about the use of MiqroForge v1 through a few steps of this example construction:

\begin{enumerate}
    \item \textbf{Select the required node}. From the list of nodes on the left, select the node you want to perform the calculation on. Generally, you can identify its function by its name. When in doubt, drag the node onto the canvas and click "?" Buttons are a way to learn more about the node.
    \item \textbf{Connecting nodes}. As with other workflow platforms, two nodes are considered connected by connecting the output endpoint of one node (right) to the input endpoint of the next node (left). Therefore, the node on the far left of this process will also start running first.
    \item \textbf{Fill in the node inputs}. Each node represents a different algorithm and therefore inevitably brings some hyperparameters. We have provided descriptive text for each node (click "?" buttons), users can quickly grasp how to fill in nodes by reading these instructions.
    \item \textbf{Run and check the results}. In the Run menu on the right, tap the Run button. These nodes are executed in a connected workflow from left to right. Some nodes have output on the Web UI, and we can see the drawn energy curve picture. Other data is saved, which is mentioned in the subsequent data section.
\end{enumerate}

More ways to use the user interface will be described in detail in the documentation.

\section{Node}

In MiqroForge, a node is defined as a single executable unit. This is similar to other (non-scientifically computed) workflow platforms, with a slight difference from the management of cross-scale simulations. Often, in cross-scale computing, researchers want to get things done in as many steps as possible. In contrast, MiqroForge advocates for increased node reusability. Therefore, properly selecting a node's function will make it simpler to accomplish.


In MiqroForge v1, nodes are implemented using docker containers. Each node is deployed and run in one or more containers, which provide the dependencies, resources, and isolated environment required by the nodes, ensuring that tasks are repeatable and portable across different computing resources.


A node must implement a contract: (1) a valid node.json; (2) an executable that reads a machine‑generated config.json indicated by \code{${input_config_path}}; (3) a \code{performance_config.json} for scheduling; (4) optional examples for self‑test.

\subsection{Node File}

In a node, there are several files necessary for it to function properly. Users can modify nodes or add new nodes by learning these files. Take the \code{PySCF-HF} node, which uses the \code{Hartree-Fock} method to calculate molecular energy and orbital information, as an example, and use the following command to enter the container:


\begin{lstlisting}[language=bash]
docker exec -it PySCF-HF-Node bash
\end{lstlisting}

Normally, all files are stored in the \code{/app/} directory,

\begin{lstlisting}[language=bash]
cd /app/PySCF-HF/
\end{lstlisting}

The directory levels are as follows,\\

\dirtree{%
.1 PySCF-HF/.
.2 node.json \DTcomment{Core configuration file}.
.2 help.md \DTcomment{Node documentation}.
.2 performance\_config.json \DTcomment{Resource scheduling configuration file}.
.2 script/.
.3 main.py \DTcomment{Executable scripts}.
.2 example/.
.3 test\_config.json \DTcomment{Test case I/O variables}. 
.3 h2o-0p85.xyz \DTcomment{Test case input file}.
}

\code{node.json} is the core configuration file of the entire node, which defines the identity and behavior of the node. It contains the unique ID, name, version, and other meta information of the node, and describes the input and output interfaces of the node (including upstream computing nodes, front-end user inputs, and downstream output results). In addition, it clarifies the commands, dependencies, and necessary contact information required for node execution, and provides sample configurations. When loading nodes, the platform will prioritize parsing \code{node.json} to build the interface and connection logic of nodes in the workflow.


\code{help.md} is a node instruction manual for end users and developers, providing node function introduction, parameter description, common problems, and operation examples. The file can be rendered directly into Markdown format and integrated into the Web UI for improved usability.


\code{performance_config.json} is used to describe the performance of nodes at different task scales, which is an important basis for intelligent scheduling systems. The file records the benchmark performance data of the node, the resource demand estimation formula (such as the relationship between the number of molecular orbitals and memory and CPU), and gives reasonable parallel calculation suggestions and running environment information. When the platform allocates compute resources, it refers to the file for automatic optimization.


\code{script/main.py} is the node's execution master program that contains specific scientific computing logic. The script parses the input configuration (usually from \code{config.json}), processes molecular structure data (such as \code{.xyz} files), and calls computational frameworks like \code{PySCF} to complete quantum chemistry calculations like Hartree-Fock. Intermediate and final result files (e.g. \code{.chk} and \code{.json}) are generated during execution.


The \code{example/} folder provides complete usage examples, including configuration files, input molecular structures, expected results, etc., to help users quickly understand node usage and operation processes. The example is self-interpretable and supports the correctness of the validator function to run the validator function with one click.


\subsection{Node File Format}

\subsubsection{node.json} \label{sec:nodejson}

The main information of a node, such as input and output, node version, Web UI display specification, etc., is integrated into the \code{node.json}. 

\begin{lstlisting}[language=JSON]
{
    "id": "684bee08-a78d-4b1f-87a8-91910ca81f38",
    "name": { "cn": , // Chinese name of this node, Web UI will displays the words. If missing, it will be displayed in English.
              "en": "Hartree-Fock (PySCF)"},
    "version": "1.0.0",
  
    "input":{
        "upstream":[ 
            {"var": "struc",  // The variable name and needs to be consistent with 'main.py'
             "name": { "cn": "", 
                       "en": "structure"},
             "description": "Import molecular structure"},  // A description of the input is displayed in the Web UI when the mouse hovers over the input.
        ],
        "web":[ 
            {
                "var": "basis", 
             	"name": { "cn": "",
                       "en": "atomic basis"},
             	"description": "",
             	"ui": {"options": ["sto3g", "631g", "ccpvtz"]}}, // Input taken directly on the Web UI with options like display.
            {"var": "unit", 
             "name": { "cn": "",
                       "en": "coordinate unit"},
             "description": "",
             "ui": {"options": ["angstrom,", "Bohr"]}},
        ]
    },
    "output":{
        "downstream":[
            {
                "var": "scf_obj", 
             	"name": { "cn": "",
                       "en": "scf object"},
             	"description": ""
            },
            {
                "var": "ene", 
             	"name": { "cn": "",
                       "en": "energy"},
             	"description": ""},
        ],
         "web":[
            {"var": "ene", 
             "name": { "cn": "",
                       "en": "energy"},
             "description": "",
             "ui": {"plain_text"}} // 
        ]
        
       		
    },
    "performance_config_path": "/app/PySCF-HF/performance_config.json",
    "example_config_path": "/app/PySCF-HF/example/test_config.json",
    "contact": {
        "name": "Quantum Computational Chemistry Group, Miqro Era",
        "email": "wuchuixiong@miqroera.com",
    },
    "execution_command": "python /app/PySCF-HF/script/main.py  --config_path ${input_config_path}" // The ${input_config_path} variable must be included
} 
\end{lstlisting}

The following table is intended to analyze in detail the structure and content of the node configuration file \code{node.json} in the MiqroForge v1 system. Node configuration files are key to ensuring that individual compute nodes can be properly deployed, initialized, and functioning efficiently. This table provides users with a clear understanding of what each configuration item means, data types, default values, and whether it is required. In addition, some example values are provided to help you understand how to configure it according to your actual needs.


\subsubsubsection{Basic Information}

The filling specifications are shown in Table \ref{tab:node_basic}.

\begin{table*}[!ht]
\begin{tabularx}{\textwidth}{lcX}
\toprule
\textbf{Field Name} & \textbf{Required} & \textbf{Explanation} \\
\midrule
id & Yes & A globally unique identifier for a node, which is automatically generated when a node is initialized  \\
name & Yes & Name of this node, Web UI will displays the words.  \\
version & No &   \\
\bottomrule
\end{tabularx}
\caption{Basic information of \code{node.json}
\label{tab:node_basic}}
\end{table*}

\subsubsubsection{Inputs/Outputs Informations}

\begin{center}
\includegraphics[width=\linewidth, clip] {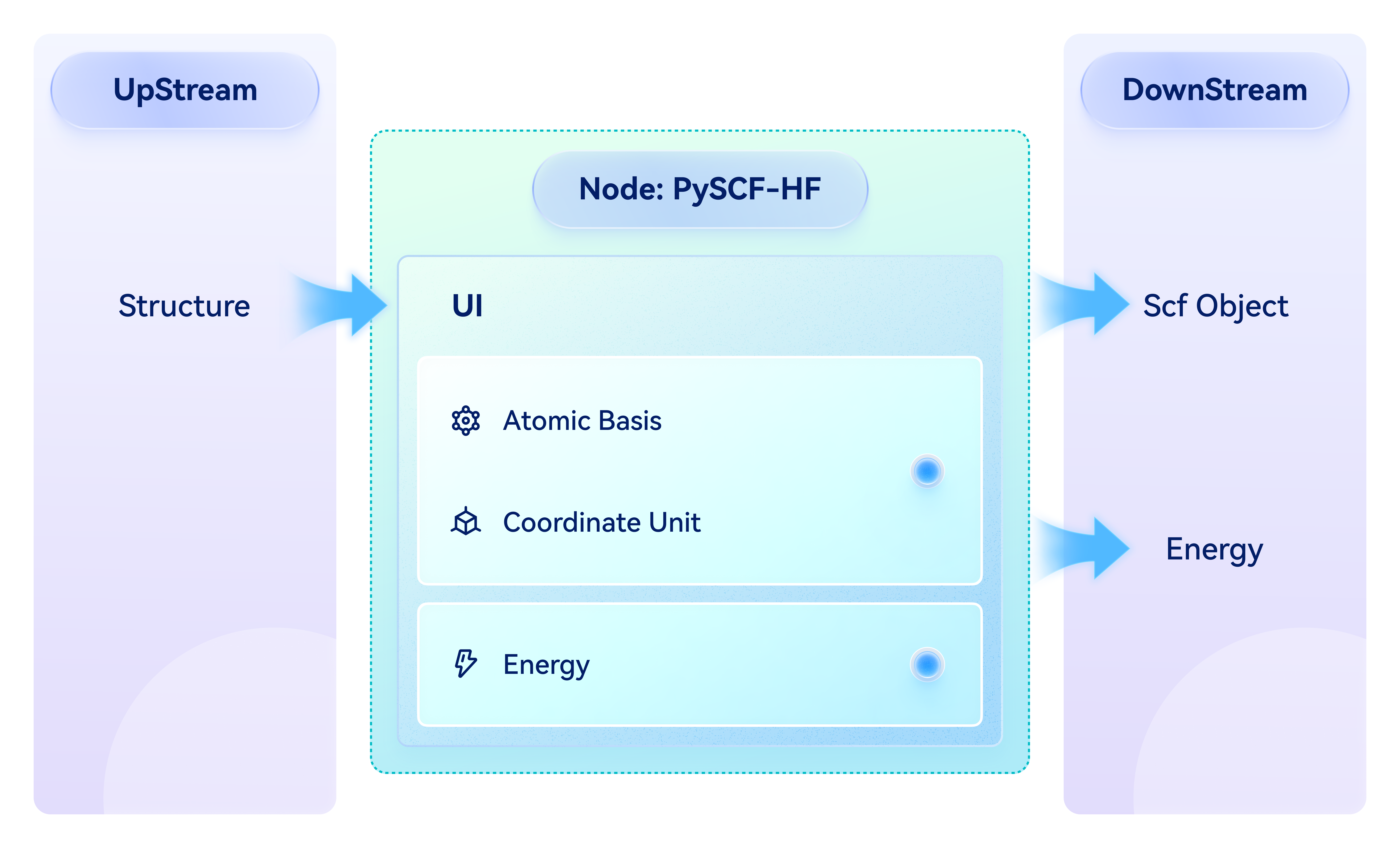}
\captionof{figure}{\small Schematic diagram of node inputs and outputs}
\label{Figure: nodeIO}
\end{center}

As shown in Figure \ref{Figure: nodeIO}, the node input has a upstream input and a Web UI input. The upstream input represents the data flowing from the predecessor nodes, while the Web UI input represents the data entered/uploaded from the Web UI. The same happens with the output. The filling specifications are shown in Table \ref{tab:node_io}.


\begin{table*}[!ht]
\begin{tabularx}{\textwidth}{lcX}
\toprule
\textbf{Field Name} & \textbf{Required} & \textbf{Explanation} \\
\midrule
var & Yes & True for \textbf{all} I/O subclasses: for \code{main.py} or other executables, this variable name is used to get input or save output.  \\
name & Yes & True for \textbf{all} I/O subclasses: Name of this I/O, Web UI will displays the words.  \\
description & No & True for \textbf{all} I/O subclasses: A description of the input is displayed in the Web UI when the mouse hovers over the input. \\
ui & Yes & True for \textbf{Web UI} I/O subclasses: Input taken directly on the Web UI with options like display.\\
\bottomrule
\end{tabularx}
\caption{I/O information of \code{node.json} 
\label{tab:node_io}}
\end{table*}

\subsubsubsection{Test and Run Informations}

For executable scripts, MiqroForge-Node requires it to read input and keywords from a specific configuration json file. The filling specifications are shown in Table \ref{tab:node_tr}. We will talk about this in detail in the next subsection.


\begin{table*}[!ht]
\begin{tabularx}{\textwidth}{lcX}
\toprule
\textbf{Field Name} & \textbf{Required} & \textbf{Explanation} \\
\midrule
performance\_config & Yes & Resource growth configuration files  \\
example\_config\_path & No & The configuration file for the test case  \\
execution\_command & Yes & The node executes a command that must contain the environment variable \code{${input_config_path}} \\
\bottomrule
\end{tabularx}
\caption{Test and run information of \code{node.json} 
\label{tab:node_tr}}
\end{table*}

\subsubsubsection{Contact Information}

The filling specifications are shown in Table \ref{tab:node_contact}.

\begin{table*}[!ht]
\begin{tabularx}{\textwidth}{lcX}
\toprule
\textbf{Field Name} & \textbf{Required} & \textbf{Explanation} \\
\midrule
name & Yes & Author name and related information  \\
email & Yes & Author's email and contact information  \\
\bottomrule
\end{tabularx}
\caption{Contact information of \code{node.json} 
\label{tab:node_contact}}
\end{table*}

\subsubsection{main.py or other executable file}

\begin{lstlisting}[language=python]
import argparse
import json
from pyscf import gto, scf

if __name__ == "__main__":
    
    parser = argparse.ArgumentParser()
    parser.add_argument(
    '-c', '--config_path', type=str,
    help='configuration file path'
	)
    f = open(
        parser.parse_args().config_path,
        'r'
    )
    conf = json.loads(f.read()) # dict from configuration file
    
    mol = gto.M(atom=conf["struc"])
    mol.basis = conf["basis"]
    mol.unit = conf["unit"]
    mol.build()
    
    mf = scf.RHF(mol)
    ene_hf = mf.run().e_tot

    mf.dump_chk(conf['scf_obj'])
    with open(conf["ene"], 'w') as f:
        f.write(str(ene_hf))
\end{lstlisting}

The \code{main.py} file first loads the molecular structure, basis and other parameters from the configuration file to build a Hartree-Fock calculation model, and then performs a self-consistent field calculation to obtain the HF energy of the system, and then writes the calculated energy results to the file, and persistently saves the current HF calculation object so that it can be directly loaded and used in subsequent processes.


For new nodes or modifying nodes, the selection and implementation of functions will not be interfered with much. MiqroForge only requires users to pass inputs and outputs to the platform through specific variables. As shown in the example above, the \code{PySCF-HF} node contains a subfield name named \code{"var"} for each input and output in the \code{"input"} and \code{"output"} field names in \code{node.json}. It can be found that \code{main.py} contains these variables in both the read and output.


The logic is this: the user first applies to MiqroForge in \code{node.json} to get a molecular coordinate file called \code{struc} from upstream. When MiqroForge runs the entire workflow, it saves the molecular coordinate file in a shared space and writes this address to a configuration json file. Then execute,


\begin{lstlisting}[language=bash]
python /app/PySCF-HF/script/main.py  --config_path ${input_config_path}
\end{lstlisting}

This command is also specified by the user. Therefore, if the code{bash} script is executed, the command can be written similarly:


\begin{lstlisting}[language=bash]
./app/PySCF-HF/script/main.sh ${input_config_path}
\end{lstlisting}

\code{${input_config_path}} here is the address of the configuration json file mentioned earlier. The user can then get the address of the molecular coordinate file from the configuration json file and load the file. In this example, this loading uses \code{python}'s \code{argparse} and \code{json} libraries.


When you want to output a variable, you only need to declare a variable name in the configuration json and output the file to the location where the variable is pointing. It is worth mentioning that all outputs need to be saved as a file. However, inputs belonging to the Web UI will be written directly to the configuration json.


\begin{table*}[!ht]
\begin{tabularx}{\textwidth}{lX}
\toprule
\textbf{I/O categories} & \textbf{variable content} \\
\midrule
upstream input & path \\
Web UI input & string\\
downstream output & path \\
Web UI output & path \\
\bottomrule
\end{tabularx}
\caption{Although the I/O variables obtained from the configuration json are all string information, some of them are direct information, while others are addresses from which the executable needs to obtain further information.}
\label{tab:io_info}
\end{table*}

\subsubsection{examples/test\_config.json}

In MiqroForge v1, each node image contains a standardized test profile \code{test_config.json}. This file is located in the node's \code{/app/example/} directory and is a key tool for validating the node's functionality. It is provided for testing in the case of a single node. We strongly recommend that users complete such examples as well when preparing their own new nodes. This document is not mandatory.


As mentioned above, in the actual operation of the node, \code{main.py} gets the I/O information from the configuration json file pointed to by \code{input_config_path} generated by MiqroForge. When writing an executable, the user needs to ensure that the \code{main.py} or other executable program reads the variable name from the configuration json file. These variable names should be consistent with the variable names in \code{node.json}.


\begin{lstlisting}[language=JSON]
{
    "basis": "6-31g", 		
    "unit": "Angstrom",	 	
    "struc": "/app/pyscf-hf/example/h2o-0p85.xyz",	
    "scf_obj": "/app/pyscf-hf/example/h2o-0p85-mol.chk",
    "ene": "/app/pyscf-hf/example/h2o-0p85-hf-energy.json"
}
\end{lstlisting}

Once you have a separate test case, run a single-node test in the installation environment using the following command:


\begin{lstlisting}[language=bash]
docker exec <container_name> python /app/script/main.py \
  --config_path /app/example/test_config.json
\end{lstlisting}

\subsubsection{performance\_config.json} \label{sec: p_c_json}

MiqroForge sets a basic resource usage for each node, which is 4 cores and 1 GB of memory. This value can be modified in the base configuration file of MiqroForge. For individual nodes, \code{performance_config.json} is used to provide node resource scheduling information. When your node only needs a fixed resource value, fill in it as follows:


\begin{lstlisting}[language=json]
{
      "recommend_min_config": "cpu: 4, memory MB: 600",
}
\end{lstlisting}

Furthermore, MiqroForge uses AI Agent to intelligently schedule node resource usage, greatly accelerating the overall workflow computing speed. This section is detailed in section VI of this article.


\subsection{Add a New Node}

Once the above files are ready, a new MiqroForge node can be created with just a few additional commands\footnote{Ensure Docker is installed and running before proceeding.}. Once the node is successfully created, you can find it in the Web UI and use it to build your own workflows.


\begin{lstlisting}[language=bash]
docker pull harbor.cl.inside/miqroforge/node-base:latest
# Pull the base image
docker run -d --name node_temp miqroforge/node-base tail -f /dev/null
# Create a test container
docker exec -it node_temp /bin/bash
# Enter the test container

<...>
# Configure the environment and prepare the necessary documents
exit

docker exec node_temp python /app/script/main.py \
  --config_path /app/example/test_config.json
docker commit node_temp your_image_name:tag
# Test the container and submit the image
miqroforge --addnode your_image_name:tag
# Commit node
\end{lstlisting}

\section{Input/Output Information}

Information serves as the core content carrier in the flow, facilitating interaction between nodes and embodying scientific logic. It includes tangible data such as molecular structures and energy values, as well as derived knowledge like computational processes and quantum state distributions. Through standardized classification and flow, information acts as an invisible link connecting various stages of quantum chemistry research.

In cross-scale platforms, information classification is the core supporting element of workflow, which directly affects the collaborative efficiency of multi-scale simulation. This section elaborates on the classification logic in workflow: by reasonably merging information types, standardized processing of data and nodes can be achieved, thereby improving information processing efficiency and system manageability. The classification mechanism not only adapts to the requirements of classical-quantum hybrid computing, but also dynamically responds to data conversion between simulations at different scales. In addition, the classification system optimizes the visual presentation of information, which echoes the intuitive visual interface of MiqroForge and helps to display information more clearly in the interface, while ensuring the accuracy and reliability of information. This ensures that the flow of information conforms to the logical order of the workflow and meets the diverse research needs of information reuse and sharing. This is of great significance for MiqroForge's data repository to achieve efficient data sharing, and injects momentum into the research progress in the field of quantum chemistry.

\subsection{Information Standards}

To achieve efficient flow and collaborative utilization of information in cross-scale research, this study constructed a classification framework covering the entire process data, dividing information into two main categories: "Naturally (N-class)" and "Computational (C-class)", and further subdividing them into secondary and tertiary subcategories, covering a complete information spectrum from raw experimental records to complex computational results.

"Naturally" (N-class) information is a dataset formed through standardized digital modeling with strict physical definitions, focusing on the inherent properties of the system, with clear physical units and standardized formats. In MiqroForge, we categorize it into three types:

\begin{enumerate}
    \item Structure: This type of information is further divided into subcategories of Molecular and Crystal. Atomic coordinates, element types, and other information are stored in .xyz or .chk files, with $Å$ as the standard unit. Structural parameters such as bond length and bond angle can be directly derived from the coordinates;
    \item Energy: Record the intrinsic energy state of the system in the form of "numerical and unit". The numerical value is float, and the unit is a.u. or Hartree. If there are other units, such as $KJ/mol$, $eV$, etc., they will be automatically converted to a.u. or Hartree to ensure consistency of information;
    \item Electronic density: Provides information on the electronic structure of the system and focuses on its microscopic distribution characteristics. By storing grid coordinates and electron density values in a. cube file, the electron density $\rho(r)$ using the default unit of $e/Bohr^3$.
\end{enumerate}

"Computational" (C-class) information is a dataset generated through the processing of computational software or algorithms, which typically relies on specific software/package and has specific formatting requirements. Currently, it is divided into algorithms and some visual outputs. Algorithms include HF/post-HF class and Quantum Computation class:

\begin{enumerate}
    \item HF/post-HF: Calculation results and core parameters based on the Hartree-Fock (HF) theoretical framework and subsequent high-precision corrections. Under this algorithm, a series of computational objects will be generated, which we further classify: 
    \begin{itemize}
        \item SCF: Dependent on PySCF software package, stored as .chk file.
        \item mol data: Generated through the ffsim software package and stored as a .chk file.
        \item ccsd.t2: Dependent on PySCF software package, stored as .chk file.
    \end{itemize}
    \item Quantum Computation: Based on quantum computing hardware/simulators, simulate quantum states, reaction pathways, etc.
    \begin{itemize}
        \item Quantum Circuit(QC): the format specification is QASM text.
        \item QC Measurement Result: after measuring the circuit using a quantum computer, generate information results in the format of a dictionary.
    \end{itemize}
    \item Figure: image data generated based on computational derivation results, corresponding visualization files are generated using tools such as Matplotlib, and allowing formats like .png, .jpg, etc.
\end{enumerate}

N-class information reflects the intrinsic properties of the system, usually covering multiple core observation dimensions in secondary subclasses without further subdivision; C-class originates from computational deduction and combines the characteristics of multiple algorithm branches and single algorithm multilevel derivation. Within the same algorithm framework, multilevel data such as intermediate states, final results, and visual mappings will be generated. It usually needs to be classified into three subcategories and may even be expanded to four or more levels in the future to adapt to the increasingly complex hierarchical logic of computational deduction.

\begin{center}
\includegraphics[width=\linewidth, clip] {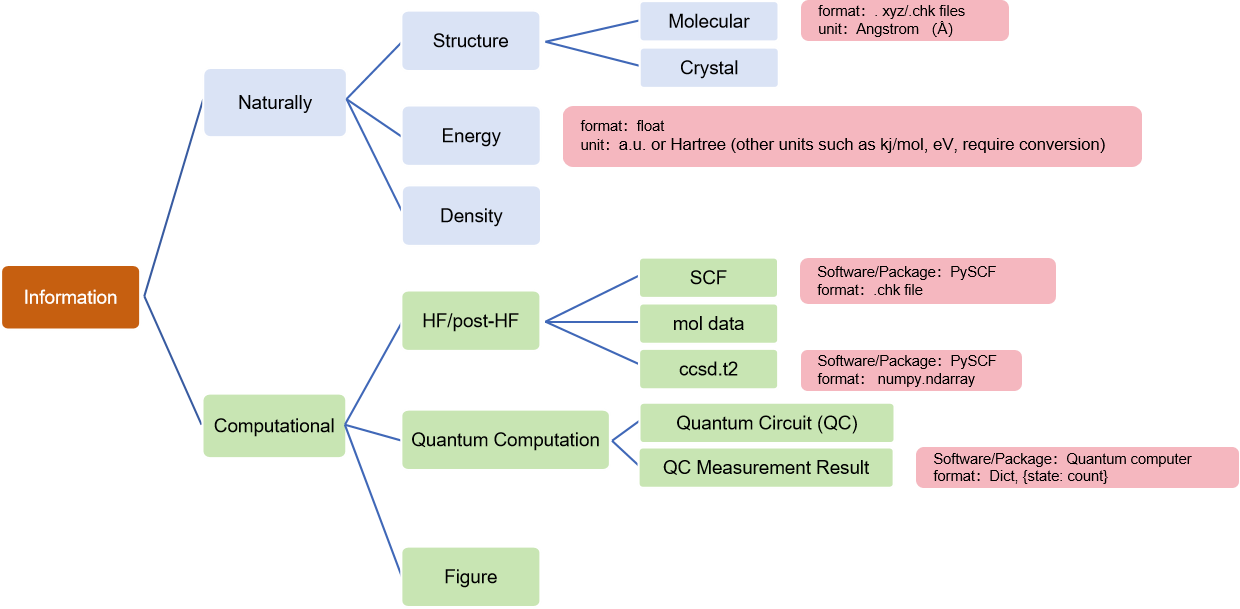}
\captionof{figure}{\small The hierarchical structure of an information classification system, where blue squares represent "Naturally", green for "Computationally" and red represents explanations of specific information.}
\label{Figure:infor}
\end{center}

Figure \ref{Figure:infor} shows the information classification and standards currently involved in MiqroForge, visually presenting the hierarchical structure and core content of the classification system, providing a reference for the standardization process and the collaborative utilization of information in cross-scale research. The content in the red text box is a standardized explanation of specific information, only showing partial information. The complete document content can be found after downloading MiqroForge.

\subsection{Information Encoding}

 To avoid confusion in information storage, we have established a unified information classification directory structure:\\
\dirtree{%
.1 MiqroForge/Info\_class/.
.2 help.md\DTcomment{Help users quickly view information classification}.
.2 N/.
.3 1.Structure/.
.4 1\_Molecular.txt \DTcomment{Describe the standard of the information}.
.4 2\_Crystal.txt.
.3 2\_Energy.txt.
.3 3\_Density.txt.
.2 C/.
.3 1\_HF\&post\_HF/.
.4 1\_SCF.txt.
.4 2\_mol\_data.txt.
.4 3\_ccsd.t2.txt.
.3 2\_Quantum\_Computation/.
.4 1\_Quantum\_Circuit.txt.
.4 2\_QC\_Measurement\_Result.txt.
.3 3\_Figure.txt.
}

There is a \code{help.md} file in the directory of \code{MiqroForge/Info_class/}, which helps users quickly understand the information standards for all categories. In addition, it contains two subcategory folders, \code{N/} and \code{C/}, representing the primary classification of information. In addition, these subcategories are further subdivided into multiple subdirectories to organize specific types of information. Txt files, like \code{2_Energy.txt}, provide detailed specifications and standards for each type of information, including format, units, and other content. Here is the file \code{1_Molecular.txt}:


\begin{lstlisting}[language=PlainText]
Info_id:
N_1_1.A.xyz
N_1_1.A.chk

Primary_Classification Number: N
Secondary_Classification Number: 1
Tertiary_Classification Number: 1
Unit: A
File Format: .xyz/.chk
\end{lstlisting}

The file provides the "info\_id" of the information, which serves as the identification number for the information, and is filled in the \code{node.json} file according to the established format: \code{"N/C_num1_num2.unit/software.format"}, \code{"N/C"} as a fixed prefix, representing the primary classification to which the information belongs, and \code{"num1/num2"} respectively represent the corresponding numbers of the secondary/tertiary classification. As mentioned earlier, with the development of the platform, there may also be numbers such as \code{"num3"} and \code{"num4"}. \code{"unit/software"} and \code{"format"} respectively represent the unit/software and format of information. This design integrates classification logic into numbering, which not only intuitively reflects the hierarchical attribution of information, but also ensures the uniqueness of each piece of information, thereby effectively improving the efficiency of information recognition and management. Here is an example of the information section in \code{node.json}:

\begin{lstlisting}[language=json, breaklines=true]
"input":{
        "upstream":[ 
            {"name": "structure",
             "description": "molecular structure",
             "info_id": "N_1_1.A.xyz"}, 
        ],
    }
"output":{
        "downstream":[
            {"name": "scf object",
             "description": "Calculation object from pyscf internal structure",
             "info_id": "C_1_1.pyscf.chk"},
        ],
    }
\end{lstlisting}

By encoding information, MiqroForge has established a standardized I/O exchange mechanism. When the "info\_id" of the input is consistent with that of the upstream output, it can ensure the smooth transmission of information. If matches incorrectly, such as when a node mistakenly input \code{"C_1_1.pyscf.chk"} to call \code{"N_1_1.A.xyz"}, the system will throw a prompt saying "Information encoding does not match, expected input N\_1\_1.A.xyz, but actually receives C\_1\_1.pyscf.chk and cannot complete information transmission". This standardized design can effectively eliminate process interruptions caused by inconsistent information formats, allowing users to focus on algorithm development and scientific problem exploration, thereby promoting the intelligent collaborative research development of MiqroForge platform in fields such as computational chemistry and material simulation, and significantly improving research efficiency.

\section{Intelligence System}

The hybrid pipelines of MiqroForge model complex scientific workflows as directed acyclic graphs (DAGs), where each node represents a specialized computation, such as quantum circuits, chemistry kernels, biological simulations or materials science analyses, and edges define data and dependency flows. An intelligent scheduler evaluates the resource requirements of each step in real time, deciding whether to pause execution until resources become available or to dynamically provision capacity and launch the next task.

Stage-aware scheduling reveals that naïve, blocking-style execution on standalone systems or modest cloud instances—typical setups for our target users—can take more than twice as long as an intelligently orchestrated run. Automated, agent-driven scheduling not only improves throughput and resource utilization but also eliminates the need for researchers to manually tune runtime parameters for each DAG branch.

While platforms such as NVIDIA’s DGX Quantum and Microsoft’s Discovery showcase the potential of tightly integrated AI–QC–HPC systems, they typically require large shared clusters and often demand significant modifications to user code and infrastructure. These constraints make them inaccessible to many domain scientists who work with single-node systems or small-scale cloud environments.

MiqroForge addresses this gap by introducing a non-intrusive, agent-based orchestration engine that operates on top of existing DAG definitions. Our agents—powered by foundation large language models (LLMs) and retrieval-augmented generation—continuously learn from resource configuration files and scheduling policies, dynamically pausing or initiating tasks without altering user scripts or requiring specialized hardware. This approach preserves both performance and flexibility, making intelligent compute scheduling accessible and practical for every researcher.

\subsection{Principle of Integration}

MiqroForge employs AI agents as dynamical scheduler and optimizes DAG formatted workflows. The scheduling process is guided by a semantic query mechanism, in which the AI agent retrievals DAG nodes and their runtime parameters using a predefined vocabulary of task descriptions and resource annotations.

At the beginning of a DAG execution, the system performs an initial semantic query, during which the AI  agent analyzes the full structure of the workflow. This includes identifying computational characteristics of each node, estimating resource requirements, and available hardware. Based on this global view, the agent generates a preliminary execution plan that prioritizes resource-efficient scheduling.

During execution, when each node completes, the system triggers an interim scheduling query before initiating the neighbor node. Dynamic queries enable the agent to reassess the current resource state, including:

\begin{itemize}
    \item Current available resources and total resources
    \item Ongoing task executions and their estimated completion times
    \item The resource intensity of the upcoming node
\end{itemize}

If the upcoming node is identified as a high-resource-consumption task(\code{include=True} in \code{performance_config.json}), the agent treats whether to :

\begin{itemize}
    \item Wait for currently executing tasks to complete for consolidating resources
    \item Proceed with immediate launching, if sufficient resources are available without causing contention
\end{itemize}

For nodes that are part of a dependency chain with potential blocking behavior, the agent will perform a look-ahead analysis during the queries. This involves roughly estimating the cumulative execution time of the entire subgraph, allowing the agent to make decisions that minimize overall workflow execution time while balancing resource utilization.

Each execution and scheduling are recorded in an internal database. The AI agent continuously refines its scheduling strategy through adaptive learning, leveraging foundation models enhanced with retrieval-augmented generation (RAG) with execution records.

The integration method the agent applied is the attached DAG node with assigned command and resources. MiqroForge separates the execution from the agent, though nowadays the ability of function call becomes popular, and no requiring modifications to user-defined DAGs or execution scripts.

In addition, there is a frontend-related AI integration named recommendation. With the documentations and example tutorials, the agent has the basis knowledge of scientific workflows. With the records of running jobs, the experiences will be inherent and actively learned. The recommendation of the next node relies on the retrieved subgraphs containing the current node and intelligence lying behind the fundamental model. We have to emphasize that only the standard nodes or the user's historical usage of node will be dashed out in the webpage.

\subsection{Let the AI knows}

The performance profile of a node, such as resource\_function and scalability, etc., is listed in the \code{performance_config.json}

\begin{lstlisting}[language=json]

{
      "include": true,

      "resource_function": "memory MB = 2*'nao' ('nao' < 200); memory MB = 0.5*'nao' (200 < 'nao' < 1000)",
      "scalability": "Number of orbitals is based on the chosen basis set. Memory requirments increase with the number of orbitals, but not in a strictly linear one.",
      "recommend_min_config": "cpu: 4, memory MB: 600",
      "environment": "",

      "benchmark_points": [
        {
          "molecule": "C6H6",
          "num_atoms": 12,
          "basis": "631g",
          "nao": 66,
          "cpu": 4,
          "memory MB": "112",
          "time": "0.14s"
        },
        {
          "molecule": "C6H6",
          "num_atoms": 12,
          "basis": "cc-pvtz",
          "nao": 264,
          "cpu": 4,
          "memory MB": "123",
          "time": "6.7s"
        },
      ],
}

\end{lstlisting}

MiqroForge will provide AI agents with many information. According to the argument "input"(section \ref{sec:nodejson}), the necessary vars will be retrieved directly or a pre-run for discovering implicit vars. MiqroForge takes each node's "performance\_config\_path", parses the profile, and handles them to the AI agent for next step scheduling. Thus, be careful with the performance profile, which will be the critical judgment to arrange the whole computing resources.

\begin{table*}[!ht]
\begin{tabularx}{\textwidth}{lX}
\toprule
\textbf{Field Name} & \textbf{Description} \\
\midrule
*\_resource & All types of resources, including physical capacity, current
utilized resource, and available resources. \\
resource\_function & Explicit mathematical function for input vars and resource consumption. \\
scalability & Textual description for implicit function instead of explicit one or behaviors out of the normal application domain.  \\
benchmark\_points & Collection of benchmark task cases on specific inputs. (Note: sensitive to hardware, hard to calibrate)  \\
recommend\_min\_config & If no additional notes, it defines the minimal resources. \\
\bottomrule
\end{tabularx}
\caption{Examples of essential information for scheduling
\label{tab:scheduling_info}}
\end{table*}

The intelligence system runs on the key fields in the MiqroForge internally. As shown in the Table \ref{tab:scheduling_info}, several fields in the \code{performance_config.json} for optimizing resource allocation. A hidden *\_resource field captures comprehensive hardware resource information within the system, including physical capacity, currently utilized resources, and constraints on available resource quantities. To model resource demands more precisely, the resource\_function field defines the explicit mathematical relationship between input variables and resource requirements (e.g., CPU, memory), enabling predictive allocation based on workload characteristics. Complementing this, the scalability field provides a textual description of how resource needs scale with input size, particularly useful when explicit functions are unavailable or when behavior deviates outside defined domains. The benchmark\_point field records empirical data from user-conducted tests, detailing specific input parameters and their corresponding execution times. Although the performance is hardware-dependent and results shall be applied case by case, these benchmarks help calibrate performance expectations actively in long-term usage. Finally, the recommend\_min\_config field suggests typical resource allocations—specifically, the minimal resource consumption recommended for standard single-machine hardware under average conditions. However, this recommendation should always be interpreted in conjunction with the scalability and resource\_function fields, as increasing resources does not always yield proportional performance gains due to diminishing returns or system bottlenecks.

In case of using other LLM models, MiqroForge will provide the custom API in the future.

This chapter demonstrates the logic beyond the system and accordingly example fields. No interactions nor manual settings being accessible to the users.

\section{Resource Scheduling and Data Governance}

The core philosophy of MiqroForge is to abstract complex cross-scale computations into manageable workflows while ensuring task reliability, reproducibility, and security through intelligent resource scheduling and data governance. This chapter systematically explains how the platform integrates heterogeneous computing resources, implements dynamic task scheduling, and establishes a full lifecycle data management system.


\subsection{Workflow Engine}

The workflow engine serves as the carrier for scientific intent. Through the Web UI, users construct computational blueprints centered on Directed Acyclic Graphs (DAG). Each node represents an node (e.g., quantum chemistry simulation, molecular dynamics optimization), with nodes communicating through standardized JSON Schema interfaces to ensure seamless cross-scale solver collaboration. Users simply declare node resource requirements (e.g., GPU type, memory capacity, number of qubits), and the platform automatically resolves dependencies to generate execution paths. This design significantly reduces complexity in multi-step computations—for example in materials screening, outputs from first-principles calculation nodes automatically transform into input parameters for molecular dynamics simulations, forming closed-loop research processes.


\subsection{Heterogeneous Resource Pool}

To support diverse cross-scale computation demands, the platform integrates multimodal resources through an intelligent abstraction layer:

\begin{itemize}
    \item \textbf{Classical Computing Resources}: Fully compatible with x86/ARM architecture CPU clusters, supporting multi-core concurrent tasks; GPU resource pools achieve load balancing through dynamic scheduling to accelerate AI training and scientific computing.
    \item \textbf{Quantum Computing Resources}: Provide unified access to real quantum hardware and high-performance simulators, enabling precise control of quantum gate operation sequences.
    \item \textbf{Memory Systems}: Offer 16GB-1TB configurable high-speed storage to optimize data-intensive task processing.
    \item \textbf{Open Extensibility}: Adheres to OpenAPI 3.0 specifications, supporting third-party resource integration and cross-platform task distribution to break computational silos.
\end{itemize}
Experiments show this architecture improves resource utilization in hybrid quantum-classical workflows by over 40\%.


\subsection{Scheduling System}

\begin{center}
\includegraphics[width=\linewidth, clip] {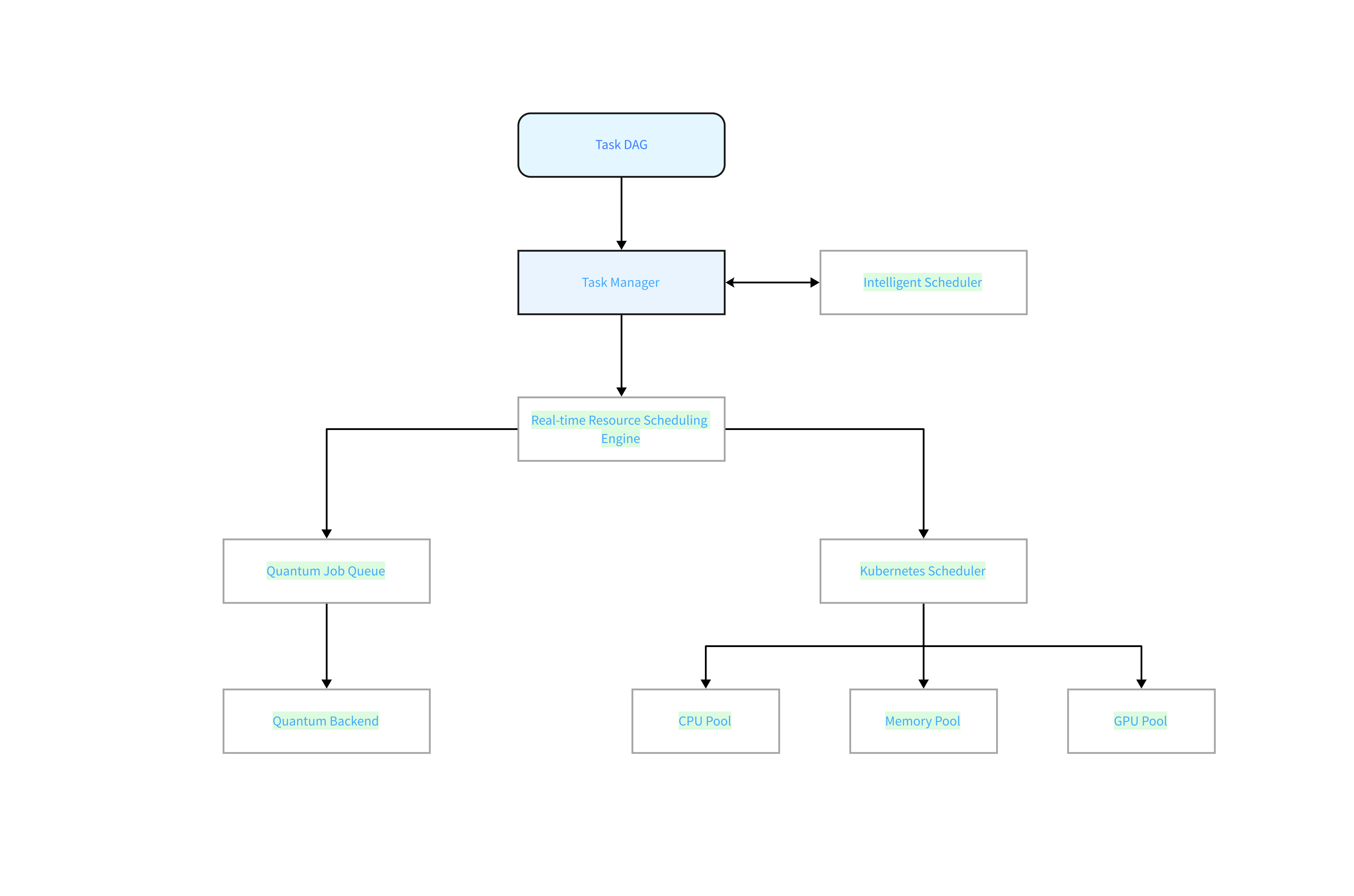}
\captionof{figure}{\small The workflow management architecture includes three layers: task management, intelligent decision-making, and task execution.}
\label{Figure: scheduling}
\end{center}

The scheduling system functions as the platform's neural center through three operational layers:

\textbf{Task Management Layer} handles full workflow lifecycle management. Upon DAG submission, the system automatically generates task instance trees, continuously monitors states, and handles exceptions (e.g., automatic node failure retries). 

\textbf{Intelligent Decision Layer} dynamically plans optimal resource allocation based on real-time cluster load and task priorities. For example, in virtual drug screening, high-throughput molecular docking tasks prioritize GPU allocation, while subsequent free energy perturbation calculations route to CPU nodes with high-precision math libraries. 

\textbf{Execution Engine Layer} performs fine-grained resource allocation:
\begin{itemize}
    \item Implements \textbf{dynamic reservation} retaining 5\% CPU resources for core platform services (logging, monitoring)
    \item Activates \textbf{CPU-exclusive policy} (\texttt{cpuManagerPolicy: static}) for latency-sensitive tasks (e.g., quantum circuit compilation)
    \item Enforces \textbf{minimum resource guarantees} (CPU $\geq$ 1 core, memory $\geq$ 1GB) for baseline execution environments
\end{itemize}

Quantum tasks are managed through dedicated queues, where special requirements (e.g., cryogenic maintenance, error correction) translate into scheduling constraints to ensure efficient classical-quantum coordination.


\subsection{Data Governance Framework}

The platform treats data as core assets through a multi-tiered governance system: 

\textbf{Workflow Version Control} captures complete execution context via snapshots, including node versions, parameter configurations, and resource consumption records. Researchers can revisit historical experiments—e.g., when reproducing material simulations from three years prior, the system automatically restores original quantum solvers and parallel computing parameters, eliminating reproducibility challenges. 


\textbf{Task Data Lifecycle} implements hierarchical management:

During execution, all intermediate data (e.g., molecular conformation trajectories) reside in high-speed storage with directory structure, this structure is persisted in the platform's data repository for easy querying: 

\dirtree{%
.1 <task\_id>/.
.2 node-<node\_id>/.
.3 job-<job\_id>/.
.4 input\_data.
.4 temp\_data.
.4 result\_data.
}

Post-execution \textbf{intelligent refinement}: Full data retained for 30 days by default; users permanently preserve critical results (free energy surfaces, quantum state fidelity) while automatically purging non-essential data. This reduces storage costs by 60\% while ensuring reproducibility.


\textbf{Data Value Extraction} utilizes visualization and analysis tools: Supports 3D molecular orbital rendering, dynamic potential energy surface displays, and cross-task data comparisons (e.g., reaction path energy barriers of different catalysts).

\textbf{Security System} integrates zero-trust principles:
\begin{itemize}
    \item Role-based granular permissions (researchers view process data, project leads manage sensitive parameters)
    \item Operation audit logs with millisecond precision
\end{itemize}


\section{Availability and Future Work}\label{sec:roadmap}

We outline near‑term milestones to improve the usability and coverage of MiqroForge while keeping the core architecture stable.

\begin{itemize}
    \item \textbf{August}: Foundation platform, the quantum sampling method (QSCI), and small‑molecule compute nodes.
    \item \textbf{September}: Intelligent compute scheduling, addition of front‑end nodes, and plotting/visualization nodes.
    \item \textbf{October}: Intelligent scheduling for hidden functions and the data report generation system; parallel multi‑task execution.
    \item \textbf{November}: Practical domain nodes, such as ionic liquids and electrocatalyst systems.
    \item \textbf{December}: Public node platform enabling users to contribute their own nodes.
\end{itemize}

These milestones are intended to make the platform incrementally more productive for both algorithm developers and applied researchers without altering the underlying node–workflow–data–AI contract.

\section*{Author Contributions}

J. Wang: Input/Output Information Standard, Official Node Production and Testing. W. Guo: Node standard, backend commands, official node test. X. Yue: Web UI, Online Document Publishing. M. Xu: Intelligent System. Y. Zheng: Mesosphere. J. Dong: Installation process, middle layer. J. Hu: Visual Element Design. J. Xia: formal analysis. C. Wu: conceptualization, investigation, supervision, writing – review \& editing.


\newpage
\printbibliography

\newpage

\renewcommand\theequation{\Alph{section}\arabic{equation}} 
\counterwithin*{equation}{section} 
\renewcommand\thefigure{\Alph{section}\arabic{figure}} 
\counterwithin*{figure}{section} 
\renewcommand\thetable{\Alph{section}\arabic{table}} 
\counterwithin*{table}{section} 


\section*{Appendix:}

We provide a CLA for contributors to ensure that community contributions remain compatible with the dual-licensing model. The project’s LICENSE and CONTRIBUTING files describe the exact terms.


\end{document}